# Intermediate Statistics, Parastatistics, Fractionary Statistics and Gentilionic Statistics.

**M. Cattani**[(*)] **and J. M. F. Bassalo** [(**)]

[(*)]Instituto de Física da Universidade de São Paulo. C.P. 66318, CEP 05315 - 970 São Paulo, SP, Brasil. E-mail < mcattani@if.usp.br>

(**)Fundação Minerva,R.Serzedelo Correa 347,1601-CEP 66035-400 Belém-Pará -Brasil

E-mail <bassalo@amazon.com.br>

## Abstract
We made in this paper a brief analysis of the following statistics: Intermediate Statistics, Parastatistics, Fractionary Statistics and Gentilionic Statistics that predict the existence of particles which are different from bosons, fermions and maxwellons. We shown the fundamental hypothesis assumed in each one of the above mentioned statistics and their main predictions and compared them with experimental results. Taking into account the works done about these statistics we could say that there is a tendency to believe that real particles, that is, those that can be observed freely, can be only bosons, fermions and maxwellons and that all other particles, different from these would be quasiparticles. Up to date in 3-dim systems only bosons, fermions and maxwellons have been detected freely. Recently in 2-dim systems have been detected the quasiparticles named anyons that have fractionary charges and spins.

## (1)Introduction.

Many works have been published about statistics that are different from the traditional bosonic, fermionic and maxwellonic. We do not intend to perform an extensive analysis about this theme but only to do a brief review of the following statistics: *Intermediate*, *Parastatistics*, *Fractionary and Gentileonic.* Since in these works we study systems composed by "particles" it is necessary to define what we understand by this word. So, we will use the name "particles" to design generically the "real particles" (that can be observed freely) and the "quasiparticles" ("collective excitations"). The particles can have internal structures like, for instance, helium atoms that are formed by a nucleus and two electrons, baryons that are formed by quarks and, as will be seen in Sec.3, *anyons* named "composed particles". Particles with no internal structure like, for instance, photons and electrons, will be called "elementary" or "punctual" particles. If in a given system the particles have internal structures but the effects of these structures can be neglected they will be taken as "elementary".

The *Intermediate Statistics* was proposed around 1940 by G. Gentile Jr.[1-3] in a thermodynamic context assuming that the particles of the system assume quantum energy levels. Gentile obtained a general expression for the energy distribution of the particles. From this general expression he obtained the particular cases of the known distributions:

bosonic, fermionic and maxwellonic. He showed that other particles could exist that do not obey these traditional statistics. He named these new particles, different from *bosons, fermions* and *maxwellons*, by "*intermediate*" particles.

The *Parastatistics* proposed in 1952 by H.Green [4] was deduced using a quantum mechanical field theory (QFT). According to their predictions would exist new particles called *paraparticles* different from *bosons* and *fermions*. The *paraparticles* with half-integer spins were named *parafermions* and those with integer spins, *parabosons*.

In 1982 F. Wilczek [5-7] created the *Fractionary Statistics*. He showed that in 2-dim systems it could exist particles with *fractionary spins*. These particles that have been called *anyons* are *quasiparticles*.

Around 1984 Cattani and Fernandes [8-15] proposed the *Gentileonic Statistics*. Their predictions were obtained within the framework of the non-relativistic Quantum Mechanics using the indistinguishability principle of identical particles in Quantum Mechanics and the *permutation group* or symmetric group $S_N$ theory. They showed that in 3-dim systems it could exist bosons, fermions and other different particles that they named *gentileons*.

In Sec.5 we present a Summary of the main results of the preceding chapters and in Sec. 6 we will see the Conclusions and Discussions.

**(1)Intermediate Statistics.**

Gentile published some papers [1-3] proposing a Quantum Statistics for 3-dim systems that would be intermediate between the Fermi-Dirac and Bose-Einstein. In other words, a statistics where the maximum occupation number of a level of energy would be given by a finite number d, different from d =1 in the case of Fermi-Dirac and d = ∞ in the case of Bose-Einstein. According to Gentile, the number d could assume any integer value between 1 and ∞. Gentile called these new particles, different from bosons and fermions, *intermediate particles*.

To calculate the new statistical distribution Gentile adopted the combinatorial method proposed by Bose[1] which is a generalization of the combinatorial method usually used in text books of Statistical Mechanics and Thermodynamics [16-19] to deduce the distributions of Fermi-Dirac, Bose-Einstein and Maxwell-Boltzmann. He assumed that the system is formed by a very large number N of identical particles with mass m, without spin and "weakly interacting". He designed by $N_s$ the total number of particles the occupy the energy level $\varepsilon_s$ and by $p_0^s$, $p_1^s$,....., $p_d^s$ the numbers of cells in the quantum phase space occupied by 0,1, ..., d particles with energy $\varepsilon_s$, respectively. In this way he got

$$N_s = \sum_{r=0}^{d} r\, p_r^s \qquad (1.1).$$

The total number of cells $Q_s$ that are available to the energy level $\varepsilon_s$ is given by

$$Q_s = \sum_{r=0}^{d} p_r^s \qquad (1.2).$$

In this way the total number of possibilities to distribute the particles in all energy levels $E_s$ would be given by the number W of possible partitions:

$$W = \prod_s (Q_s / p_0^s! \, p_1^s! \ldots p_d^s!), \qquad (1.3)$$

taking into account that

$$E = \Sigma_s N_s \varepsilon_s = \Sigma_r \Sigma_s p_r^s \, r \, \varepsilon_s \quad \text{and} \quad N = \Sigma_s N_s = \Sigma_r \Sigma_s r \, p_r^s \qquad (1.4)$$

Using Eq.(1.3) he calculated ln(W) taking into account the Stirling approximation $\ln(p!) \approx p \ln(p)$, since the p values are very large:

$$\ln(W) = \Sigma_s Q_s \ln(Q_s) - \Sigma_r \Sigma_s p_r^s \ln(p_r^s) \qquad (1.5)$$

Remembering that the entropy S is defined by $S = k \ln(W)$ [16-19], where k is th Boltzmann constant, let us maximize ln(W) considering the constraints functions E and N defined by (1.4). In this way, using the Lagrange multipliers α and β we get

$$\Sigma_r \Sigma_s ( 1 + \ln(p_r^s) + \alpha r + \beta r \varepsilon_s ) \, \delta p_r^s = 0, \qquad (1.6)$$

from which we obtain,

$$p_r^s = B_s \exp[-\gamma_s r], \qquad (1.7)$$

where $\gamma_s = \alpha + \beta \varepsilon_s$ and $B_s$ is a parameter to be determined using (1.2). Thus, substituting (1.7) into (1.2) results,

$$Q_s = \sum_{r=0}^{d} p_r^s = B_s \sum_{r=0}^{d} \exp[-\gamma_s r] = B_s (1 - \exp[-(d+1)\gamma_s]) / (1 - \exp[-\gamma_s]),$$

from which we obtain,

$$B_s = Q_s (1 - \exp[-\gamma_s]) / (1 - \exp[-(d+1)\gamma_s]) \qquad (1.8)$$

Substituting the function $p_r^s$ defined by (1.7) into (1.1) we get,

$$N_s = \sum_{r=0}^{d} r \, p_r^s = Q_s \{ 1/(\exp[\gamma_s] - 1) - (d+1)/(\exp[(d+1)\gamma_s] - 1) \}, \qquad (1.9)$$

which is the fundamental equation of the Intermediate Statistics that contains the particular cases of Fermi-Dirac, Bose-Einstein and Maxwell-Boltzmann, as will be shown below. Before to do this, taking into account that the entropy $S = k \ln(W)$ and using (1.5), (1.7) and (1.8) we see that

$$S = k \{ N\alpha + U\beta + \Sigma s \, Q_s \ln [(1 - \exp[-\gamma_s]) / (1 - \exp[-(d+1)\gamma_s])] \} \qquad (1.10),$$

where U = E is the internal energy of the system. As we must have [16-19] $\partial S/\partial U = 1/T$, from (1.10) we verify that $\beta = 1/kT$. Other relation that can be obtained from (1.10) and the Thermodynamics [16-19] is that $\alpha = -\mu/kT$, where $\mu$ is the chemical potential. In this way we see that the parameter $\gamma_s = \alpha + \beta \varepsilon_s$ can be written as $\gamma_s = (-\mu + \varepsilon_s)/kT$. At this point it is important to note that, according to Gentile hypothesis, the number d can assume the values $d = 1, 2,...,N$ independently of the number $N \gg 1$ of particles of the system.

Now, let us obtaing the limiting cases of the fundamental equation (1.9) of the *Intermediate Statistics*:

(1.a) Fermi-Dirac Statistics (d = 1).

In this statistics it is assumed that the maximum occupation number d of a given state is d =1 independently of the total number N of particles of the system. In this case using (1.9) we have,

$$N_s = Q_s \{1/(\exp[\gamma_s] - 1) - 2/(\exp[2\gamma_s] - 1)\} = Q_s /\{ \exp[(\varepsilon_s - \mu)/kT] + 1 \} \quad (1.11)$$

(1.b) Bose-Einstein Statistics (d = $\infty$).

Here it is assumed that any state can be occupied by all particles of the system, that is, d = N. Since $N \gg 1$ we put $d = N \rightarrow \infty$ into (1.9) obtaining:

$$N_s = Q_s /\{ \exp[(\varepsilon_s - \mu)/kT] - 1 \} \quad (1.12)$$

(1.c) Maxwell-Boltzmann Statistics ($\varepsilon_s - \mu \gg kT$).

In this case from (1.9) we get,

$$N_s = Q_s / \exp[(\varepsilon_s - \mu)/kT] \quad (1.13).$$

Two different approaches recently found in the literature obtaining the same results predicted by Gentile are due to Ben-Abraham [20] and to Vieira e Tsallis [21]. The titles of the papers are, respectively, "Paragas" and "D - Dimensional Ideal Gas in Parastatistics: Thermodynamic Properties". Note that in the Intermediate Statistics are taken into account only arguments of Statistical Mechanics and Thermodynamics that are valid only for systems with a very large number of particles. In these works are not analyzed the implications of the predictions with the foundations of the Quantum Mechanics [22-24], with the Quantum Field Theory (QFT) or with the Gauge Theory. In next sections we will see how these fundamental aspects are taken into account analyzing the different proposed statistics. In the thermodynamic context there was no worry to consider the spins of the particles. The analysis between the spin and statistics in the QFT context is named "spin-statistics connection".

**(2) Parastatistics.**

In the framework of the QFT it was demonstrated a famous theorem named *Pauli Theorem* [25-28] that says: "*bosons* are particles with integer spins and their creation and destruction operators $a_k^*$ and $a_k$, respectively, must obey bilinear commutation relations" and "*fermions* are particles with half-integer spins and their creation and destruction operators $a_k^*$ and $a_k$, respectively, must obey bilinear anti-commutation relations". So, according to this theorem only *bosons* and *fermions* could exist. The *intermediate* particles could not exist.

However, in 1953, H.S.Green[4], generalizing the QFT formalism showed that it could exist also particles with integer and half-integer spins, that are different from *bosons* and fermions. He named these particles *parabosons* and *parafermions* and the new statistics was called *Parastatistics*.

Following Green [4], any field Ψ(x) in the interaction representation can be expanded in the form

$$\Psi(x) = \Sigma_k \{a_{k+} \varphi_{k+}(x) + a_{k-}{}^* \varphi_{k-}{}^*(x)\}, \quad (2.1)$$

where $\varphi_{k+}(x)$ and $\varphi_{k-}{}^*(x)$ are complete orthonormal set of functions which the first ones contain only positive frequencies and the second ones only negative frequencies, the operators $a_r$ ($r = k_+$ ou $k_-$) always annihilate and $a_r{}^*$ always create a particle (with positive or negative energy). We can write, for instance,

$$\varphi_{k\pm}(x) = V^{-1/2} \exp(\pm i\, p_k{}^\alpha x_\alpha)\, \Psi_{k\pm} \quad (p_k{}^4 > 0),$$

where V is the volume a given 3-dim region, $p_k{}^\alpha$ are the possible values of the 4-vector momentum-energy and, depending on the kind of the field, $\Psi_{k\pm}$ can be an scalar, a vector or a spinor conveniently normalized. He assumed that a quantization scheme is considered satisfactory if the following condition is obeyed:

$$\partial_\alpha \Psi(x) = i\, [P_\alpha, \Psi(x)] \quad (2.2),$$

where $P_\alpha$ is the energy-momentum 4-vector of the field. He also assumed that the 4-vectores $P_\alpha$ of half-integer spin and integer spin fields obey the following relations, respectively,

$$P^\alpha = \Sigma_r\, p_r{}^\alpha\, [a_r{}^*, a_r] \quad e \quad P^\alpha = \Sigma_r\, p_r{}^\alpha\, \{a_r{}^*, a_r\}. \quad (2.3).$$

Using Eqs.(2.3) he determined the necessary conditions to satisfy (2.2). So, he showed that particles represented by half-integer spin fields would have their fields quantized in such a way that an arbitrary finite number of them could occupy the same quantum state. For these particles, called *parafermions*, would be valid the following commutation relation:

$$[a_r{}^*, [a_s, a_t]] = \delta_{rs}\, a_t - \delta_{rt}\, a_s \quad, \quad [a_r{}^*, [a_s{}^*, a_t{}^*]] = 0 \quad (2.4).$$

All relations (2.4) are satisfied if it is assumed, as usually for fermions, that

$$\{a_r, a_s\} = 0 \quad, \quad \{a_r{}^*, a_s\} = \delta_{rs} \quad (2.5),$$

where {a,b} indicates the anticommutator of a and b. However, there are an infinity of possible trilinear relations that satisfy (2.4) as, for instance,

$$a_r\, a_s\, a_t + a_t\, a_s\, a_r = 0\ ,$$

$$a_r{}^*a_s\, a_t + a_t\, a_s\, a_r{}^* = \delta_{rs}\, a_t \qquad (2.6)$$

$$a_r\, a_s{}^*\, a_t + a_t\, a_s{}^*\, a_r{}^* = \delta_{rs}\, a_t + \delta_{ts}\, a_r$$

No caso de partículas representadas por campos de spin inteiro e mostrou que $a_k{}^*$ e $a_k$ obedecem às seguintes relações de comutação,

$$[a_r{}^*, \{a_s, a_t\}\,] = \delta_{rs}\, a_t \quad , \qquad [a_r, \{a_s, a_t\}\,] = 0 \qquad (2.7).$$

All relations (2.7) are satisfied is we assume, as usually for *bosons,* that

$$[a_r, a_s] = 0 \quad , \qquad [a_r{}^*, a_s] = \delta_{rs} \qquad (2.8).$$

However, as occurs with *parafermions*, there are an infinity of possible trilinear commutation relations that satisfy the Eqs.(2.8) as, for instance,

$$a_r\, a_s\, a_t - a_t\, a_s\, a_r = 0\ ,$$

$$a_r{}^*a_s\, a_t - a_t\, a_s\, a_r{}^* = \delta_{rs}\, a_t \qquad (2.9)$$

$$a_r\, a_s\, a_t{}^* - a_t{}^*\, a_s\, a_r = \delta_{rs}\, a_t - \delta_{ts}\, a_r\ .$$

These particles were named *parabosons*. According to Green's predictions an arbitrary finite number of *parabosons* or *parafermions* can occupy the same quantum state.

It is important to note that using the irreducible representations of the symmetric group $S_N$ in the Hilbert space [8-13] the bosonic $\Psi_S(x)$ and fermionic $\Psi_A(x)$ wavefunctions are 1-dim corresponding to the horizontal and vertical Young shapes, respectively. In this context [29] the fields $\Psi(x)$ given by (2.1) of the *parabosons* and of the *parafermions* are also 1-dim. As will be seen in Sec.(4) the *gentileons* are associated with intermediate Young shapes and their non-relativistic wavefunctions Y belong to Hilbert spaces with dimensions $\tau$ going from $2^2$ up to $(N-1)^2$. From these properties we see that the Green fields $\Psi(x)$ do not describe in the non-relativistic limit [29] the gentilionic states Y as ought to be expected [10].

A detailed analysis of the Parastatistics and their applications in many physical problems can be seen, for instance, in the book of Ohnuki e Kamefuchi [29] and in many recent papers [30-33]. Ohnuki e Kamefuchi [29] analyzed the possibility of quarks be *parafermions*, as proposed by Greenberg in 1964 [34]. They concluded that the quarks cinematic properties obeying the *parafermionic* statistics cannot, by themselves, explain the quark confinement and that the *paraquarks* model is incompatible [29] with the $SU(3)_c$ color gauge.
    .

**(3) Fractionary Statistics.**
In 1982 using the electromagnetic gauge theory F. Wilczek [5-7] showed that could exist, in 2-dim systems, particles with fractionary spins that he named *anyons*. These new particles that obey a statistics that was called *Fractionary Statistics* are *quasiparticles*, as will seen in what follows [35].

Following Wilczek [5-7], let us image a particle with charge q orbiting around a very long solenoid (not penetrating it) which is parallel to a z-axis. When there is no current in the solenoid the angular momentum $l_z$ of the charge is quantized given by $l_z = m\hbar$ where m = integer. If a current is slowly turned on, the charged particle will feel an electric field $\mathbf{E}(\mathbf{r}) = -(z \times \mathbf{r})(d\Phi/dt)/[2\pi(x^2+y^2)]$, according to Faraday's law, where $\Phi$ is the flux of the magnetic field through the solenoid. It will generate a change in the angular momentum given by

$$d(l_z)/dt = [\mathbf{r} \times (q\mathbf{E})]_z = -(q/2\pi c)(d\Phi/dt). \qquad (3.1).$$

So, the total angular momentum change $\Delta l_z$ will depend only on the final flux value $\Phi$, $\Delta l_z = -q\Phi/2\pi c$. In this way the quantized angular momentum value is modified in such a way that when the flux in the solenoid is $\Phi$ the angular $l_z$ is given by

$$l_z = m\hbar - q\Phi/2\pi c \qquad (3.2).$$

This same conclusion could be obtained as follows. As the angular momentum operator is defined by $l_z = -i\hbar\partial_\varphi - qA_\varphi/c$, $A_\varphi = \Phi/2\pi$ and the electron wavefunction is given by $\Psi_{nm}(r,\varphi) = \exp(im\varphi) R_n(r)$ we obtain $l_z \Psi_{nm} = (-i\hbar\partial_\varphi - qA_\varphi/c) \Psi_{nm} = (m\hbar - q\Phi/2\pi c) \Psi_{nm}$, according to (3.2).

Eliminating the vector potential outside of the solenoid by a gauge transformation, that is, putting $\mathbf{A}' = \mathbf{A} - \text{grad}(\Lambda)$ with $\Lambda = q\Phi/2\pi\hbar c$ the electron wavefunction, that obeys a Schrödinger equation, will be now be given by $\Psi_{sm}'(r,\varphi) = \exp[i(m + q\Phi/2\pi\hbar c)\varphi] R_s(r)$. In these conditions (as there is no more a vector potential $\mathbf{A}$) the angular momentum is given by the usual operator $l_z = -i\hbar\partial_\varphi$ that applied on $\Psi_{sm}'(r,\varphi)$ gives the angular momentum value $l_z = m\hbar - q\Phi/2\pi c$. In these conditions the new wave function $\Psi_{sm}'(r,\varphi)$ can be written as $\Psi_{sm}'(r,\varphi) = \exp[i(q\Phi/2\pi\hbar c)\varphi] \Psi_{sm}(r,\varphi) = \exp[i\Delta\varphi] \Psi_{sm}(r,\varphi)$, where $\Delta = q\Phi/2\pi\hbar c$.

The wavefunction $\Psi_{sm}'(r,\varphi)$ represents a particular formed by a *charge+ electromagnetic field* called *anyon*. It is not an elementary particle strictu sensu but, a "*composed particle*" (see comments at the end of this section). Note that when the *anyon* rotates around the solenoid by an angle $2\pi$ (taking r = constant) we get,

$$\Psi_{sm}'(r,\varphi+2\pi) = \exp[iq\Phi/\hbar c] \Psi_{sm}(r,\varphi+2\pi) = \exp[iq\Phi/\hbar c] \Psi_{sm}(r,\varphi) \qquad (3.3),$$

since $\Psi_{sm}(r,\varphi+2\pi) = \Psi_{sm}(r,\varphi)$. So, (3.3) shows that due to the rotation $2\pi$ around the solenoid the *anyon* statefunction changes by a factor $\exp(iq\Phi/\hbar c)$.

Let us consider now two identical *anyons* 1 and 2, that do not interact, represented by the statefunctions $\Psi_1'(r_1,\varphi_1) = \exp[i\Delta\varphi_1] \Psi_{sm}(r_1,\varphi_1)$ and $\Psi_2'(r_2,\varphi_2) = \exp[i\Delta\varphi_2] \Psi_{tm}(r_2,\varphi_2)$. The state function of the two anyons $\Psi_{12} = \Psi_1'(r_1,\varphi_1)\Psi_2'(r_2,\varphi_2)$ is given by

$$\Psi_{12} = \exp[i\Delta(\varphi_1 + \varphi_2)] \Psi_{sm}(r_1, \varphi_1) \Psi_{tm}(r_2, \varphi_2) \qquad (3.4).$$

Let us assume that at the initial state (I) the particles are located at $r_1 = r_2 = r$, $\varphi_1 = 0$ and $\varphi_2 = \pi$ :

$$\Psi_{12}{}^{(I)} = \exp[\pi i\Delta]\ \Psi_{sm}(r, 0)\ \Psi_{tm}(r, \pi). \tag{3.5}$$

Assuming now the permutation 1→2: $\Psi_{sm}(r_1, \varphi_1)\Psi_{tm}(r_2, \varphi_2) \to \Psi_{tm}(r_1, \varphi_1)\ \Psi_{sm}(r_2, \varphi_2)$ and putting $\varphi_1 = \pi$ and $\varphi_2 = 2\pi$, taking $r_1 = r_2 = r$, we will have at the final state (F):

$$\Psi_{21}{}^{(F)} = \exp[3\pi i\Delta]\ \Psi_{tm}(r, \pi)\Psi_{sm}(r, 2\pi) \tag{3.6}$$

Taking into account that $\Psi_{sm}(r, 0) = \Psi_{sm}(r, 2\pi)$ we see, using (3.4) and (3.6) that

$$\Psi_{21}{}^{(F)} = \exp[2\pi i\Delta]\ \Psi_{12}{}^{(I)}. \tag{3.7}$$

The Eq. (3.7) shows that if $\Delta = q\Phi/2\pi\hbar c = 0$ the *anyons* behave like *bosons* and for $\Delta = 1/2$ like *fermions*. However, when $\Delta$ is in the interval $0 < \Delta < 1/2$ there are an infinite species of *anyons*, different from *bosons* or *fermions*.

In the general case [27-29] indicating by s the spin of the charge that orbits the solenoid we verify, using (3.2), that the angular momentum or spin $s_a$ anyon is given by

$$s_a = m + \Delta + s \qquad (m = \text{inteiro}) \tag{3.8},$$

showing that the *anyon* spin $s_a$ can assume fractionary values, different from the integer (bosonic) or half-integer (fermionic) spins.

Recently, according to Camino et al. [35, 36], the existence of anyons ("*Laughlin particle*" with fractionary charge) was confirmed in the context of the Fractionary Quantum Hall Effect. These *fractionary* particles cannot be found asymptotically free in nature, they are *quasiparticles*. More details about this experimental confirmation are given by Lindley [37]. In 1-dim systems are predicted particles similar to *anyons* named *exclusons* [38].

The concept "*composed particles*" was used by Jain [39,36] to describe the *anyon*, that is, the system formed by a *charge + gauge field.* More accurately, we ought to say that anyon is a *composed quasiparticle*. When the charge is a *fermion* the *anyon* is called *quasifermion* and when the charge is a *boson* it is named *quasiboson*.

In spite of considerable efforts of many authors [40] up to now there is no satisfactory QFT formulation to the *anyons*. In this way it was not possible to analyze the *spin-statistics connection*. Within the quantum mechanical context, relativistic and non-relativistic, there is a satisfactory formal understanding of the behavior and properties of the *anyons*. However, from the point of view of the applications of the theory there are many difficulties because the wavefunction $\Psi$ of a system of non-interacting *anyons* without external field can be calculated exactly only for the case of two particles; for N > 2 there is no exact solution [40, 41]. As *anyons* are intermediate particles between *bosons* and *fermions* a limited number of anyons must occupy the same quantum state. However, as we do not know to calculate exactly $\Psi$ for an arbitrary number N of *anyons* we cannot determinate the number of *anyons* that must occupy a same quantum state. Due to this limitation the properties of the systems composed by N *anyons* are estimated calculating their state equation doing an interpolation between the bosonic and fermionic state equations [42].

(3.1) Anyons and Gauge Theory.

Leinaas and Myeheim in 1977 [43] were the first ones to show, using the Gauge Theory, that in 2-dim systems it could exist quasiparticles that obey a statistics varying continuously between the bosonic and fermionic. Let us see now the connection between the gauge theory and the *anyons*. Within the framework of the gauge theory [44-46] particles with internal structures carry their own "*internal spaces*" as they move along the spacetime (x). These particles are "composed particles" [39]. The internal degrees of freedom of these particles change as they move submitted to an external field. If $\Psi(x)$ is the statefunction of the particle the change of state is performed by an unitary operator $U(x)$, that is, $\Psi'(x) = U(x) \Psi(x)$. According to the non-Abelian group theory of Yang-Mills [44-46] the operator $U(x)$ is given by,

$$U(x) = \exp[-ig(\Sigma_k \theta^k(x) F_k)] \qquad (3.8)$$

where $g = q/(\hbar c)$, q is the particle charge, $k = 1,2,3,...,n$, where n is the dimension of the group of gauge G associated with a given external field called "gauge field". The operators $F_k$ satisfy the usual commutation relations $[F_i, F_j] = ic_{ijk} F_k$ where the constants $c_{ijk}$ depend on the particular group G. The parameters $\theta^k(x)$, that are functions of the spacetime x, called "*rotation angles*", represent the internal degrees of freedom of the particle. The functional dependence of $\theta^k(x)$ with x is responsible for the connection between the internal degrees of freedom with the external field for different points of the spacetime. The external field defined by $A_\mu(x)$ that connects the internal states to the spacetime x, named also as "connection operator" is defined in such a way that the matrix elements $(A_\mu)_{\alpha\beta}$, where the states α and β are internal states of the particles, are given by

$$(A_\mu)_{\alpha\beta} = \Sigma_k \partial_\mu(\theta^k) (F_k)_{\alpha\beta} \qquad (3.9).$$

The function $A_\mu(x)$ is at the same time an external field and an operator that acts in the internal space of the particle. This operator is not an observable quantity since it depends on non observable quantities, the angles $\theta^k(x)$ and the generators $F_k$ of the symmetry group.
Using these properties of the gauge theory we can show[44-46] that the wavefunction $\Psi_o(x)$ of a *composed particle* after describing a closed path submitted to an external field changes by a phase φ, according to the transformation $\Psi(x) = U(x)\Psi_o(x) = \exp(i\varphi)\Psi_o(x)$, given by

$$\varphi = g \oint A_\mu(x) dx^\mu = g \iint F_{\mu\nu}(x) dx_\mu dx_\nu \qquad (3.10),$$

where the operator $F_{\mu\nu}(x) = \partial_\mu A_\nu - \partial_\nu A_\mu - ig[A_\mu, A_\nu]$ is known as the generalized Maxwell tensor since in the particular case of the electromagnetic it becomes the Maxwell tensor.
The Eq.(3.10) expresses the generalized Aharonov-Bohm Effect [47] that must observed for any gauge field [45,46,48].

In the particular case of the electromagnetism, when the group G is the U(1) and the external field is the vector potential $A_\mu(x)$ it can be shown that [45,46]

$$\varphi = g \oint \mathbf{A}(x) d\boldsymbol{\ell} = g \oint \text{rot}(\mathbf{A}) \cdot d\mathbf{S} = g \Phi \qquad (3.11),$$

where g = e/(ℏc), e the eletron charge and Φ the flux of the magnetic field inside the solenoid, giving exactly the same phase change φ seen in (3.3).

**(4) Gentileonic Statistics.**

Around 1984 Cattani and Fernandes [8-15] proposed the *Gentileonic Statistics*. Their predictions were obtained within the framework of the non-relativistic Quantum Mechanics assuming the *Indistinguishability Principle* of identical particle and using the *permutation group theory* or symmetric group $S_N$ theory. They considered 3-dim isolated systems with constant energy E and with an arbitrary constant number N of identical *elementary particles* (free or interacting). They showed that for these systems it could exist bosons, fermions and other different particles that they named *gentileons*. It is important to note that their predictions are not valid for *composed particles* like anyons because for these particles the symmetric group $S_2$ cannot be applied [43]. They have indicated by H the non-relativistic hamiltonian operator of the system obtained, for instance, as non-relativistic limits of the Dirac or Klein-Gordon hamiltonians. In this way the energy eigenfunction Ψ of the system obeys the Schrödinger equation HΨ = EΨ. The operator H and Ψ are functions of $\mathbf{x}_1,\mathbf{s}_1,…,\mathbf{x}_N, \mathbf{s}_N$, where $\mathbf{x}_j$ and $\mathbf{s}_j$ denote the position coordinate and the spin, respectively, of the particle j. The pair $(\mathbf{x}_j,\mathbf{s}_j)$ will be indicated by a single number j ( j =1,2,…,N) named *particle configuration.* The set of all configurations generates the *configuration space* $\varepsilon^{(N)}$. So, we write simply H = H(1,2,…,N) and Ψ = Ψ(1,2,...N).

They defined by $P_i$ the "permutation operator" (i = 1,2,...N!) that generate all possible permutations of the particles in the space $\varepsilon^{(N)}$. Since the particles are identical the physical properties of the system must by invariant by permutations. The $P_i$ constitute a symmetry group $S_N$ called Permutation Group or Symmetric Group ($S_N$) of order n = N!. Due to the identity of the particles, H e Ψ, obtained permuting the particles must be physically equivalent, that is, $[P_i, H] = 0$ e $|P_i\Psi|^2 = |\Psi_i|^2 = |\Psi|^2$. This implies that the permutations are unitary transformations and that the energy E spectrum is n times degenerate. Assuming that all functions $\{\Psi_i\}_{i=1,2...,n}$ are different and orthogonal they have associated in a one-by-one correspondence an unitary operator $U(P_i)$ so that $\Psi_i = U(P_i)\Psi$. The functions $\{\Psi_i\}_{i=1,2...,n}$ form a n-dim Hilbert space $L_2(\varepsilon^{(N)})$ of all square integrable function over $\varepsilon^{(N)}$. As the eigenfunctions $\{\Psi_i\}_{i=1,2...,n}$ are ortonormal and represent states with the same energy E they can be used with help of the *Young Shapes* (or *Diagrams*) to determine the energy eigenfunctions Ψ of systems with an arbitrary number of particles. To the horizontal and vertical *Young Shapes* are associated the statefunctions of *bosons* ($\Psi_S$) and *fermions* ($\Psi_A$), respectively, given by

$$\Psi_S = (1/\sqrt{N!}) \sum_{i=1}^{N} \Psi_i \qquad e \qquad \Psi_A = (1/\sqrt{N!}) \sum_{i=1}^{N} \delta_{Pi} \Psi_i \qquad (4.1).$$

where $\delta_{Pi} = \pm 1$, if $P_i$ is an even or odd permutation, respectively.

To the intermediate (between horizontal and vertical) Young Shapes are associated the statefunctions Ψ(α) = Y(α) where (α) indicate the *partitions* of the number N. these functions are written in following way,

$$Y(\alpha) = \frac{1}{\sqrt{\tau}} \begin{pmatrix} Y_1(\alpha) \\ Y_2(\alpha) \\ \vdots \\ Y_\tau(\alpha) \end{pmatrix} \quad (4.2)$$

where $\tau$ is the dimension of the Hilbert space which corresponds to a given partition ($\alpha$). The wavefunctions Y($\alpha$) would represent particles that are different from *bosons* and *fermions* that were named *gentileons*.

To the permutations $P_i$ (i = 1,2,…, N!) are associated unitary operators U($P_i$) that are represented by unitary matrices M($P_i$) (or D($P_i$)) in $L_2(\varepsilon^{(N)})$. In the case of bosons and fermions these matrices are 1-dim, that is, $P_i \Psi_S = (+1)\Psi_S$ and $P_i \Psi_A = (\pm 1)\Psi_A$. In the gentileons case we have $P_i Y = M(P_i)Y$ where M($P_i$) are square (p x p) unitary matrices, with p = N – 1.

Taking into account the symmetry properties of the gentileonic states Y they showed that [9-15]: (a) the gentileons cannot appear as free particles and (b) the gentileonic systems cannot coalesce. These properties suggest that *gentileons* would be *quasiparticles*. Since quarks seem to be permanently confined, never being detected freely, like usual *bosons* and *fermions*, they have studied hadronic properties assuming that quarks are gentileons [9-15].

In order to show in a simple way the fundamental symmetry properties of the functions Y($\alpha$) we will analyze in what follows only systems composed by weakly interacting particles.

(4.1) *Weakly Interacting Particles.*

Let us assume that the systems are composed by weakly interacting particles forming an "ideal gas" as was done by Gentile in Sec.1. Thus, let us indicate by $\varphi_s$ and $\varepsilon_s$ the energy eigenstates and the respective energies assumed by the particles. In this case a typical energy statefunction $\Psi$ of the N particles is written as a product of the functions $\varphi_s$, that is, $\Psi = \Psi(1,2,...,N) = \varphi_n(1)\varphi_m(2)...\varphi_k(N)$ or, to simplify the notation, $\Psi = \Psi(1,2,...,N) = u(1)v(2)...w(N)$.

Now let us consider only the cases of N =2 and N =3. For these cases we have the corresponding Young shapes, shown in Fig.1:

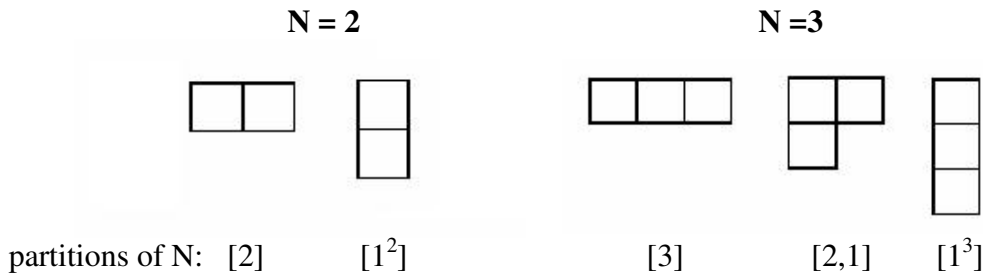

**Figure 1**. The Young shapes to N = 2 e N =3 particles with the corresponding partitions of the numbers 2 and 3 displayed bellow the respective shapes.

Since for N = 2 there is no intermediate Young shape, we can have only *bosons* and *fermions* that represented by the eigenstates $\Psi_S[2] = [u(1) v(2) + v(2) u(1)]/\sqrt{2}$ and $\Psi_A[1^2] = [u(1) v(2) - v(2) u(1)]/\sqrt{2}$. For N = 3 we have the shapes, horizontal [3], vertical [$1^3$] and intermediate [2,1]. The wavefunctions, bosonic $\Psi_S$ and fermionic $\Psi_A$ are given by

$\Psi_S[3] = [u(1)v(2)w(3) + u(1)v(3)w(2) + u(2)v(1)w(3) + u(2)v(3)w(1)$

$+ u(3)v(1)w(3) + u(3)v(2)w(1)]/\sqrt{6}$   and

$\Psi_A[1^3] = [u(1)v(2)w(3) - u(1)v(3)w(2) - u(2)v(1)w(3) + u(2)v(3)w(1)$

$+ u(3)v(1)w(2) - u(3)v(2)w(1)]/\sqrt{6}$.

The wavefunction $\Psi[2,1] = Y$ that describes the *gentileons*, associated to the intermediate shape [2,1], can represented by $Y_+$ or $Y_-$ since these functions have equal symmetry permutation properties:

$$Y_+ = \frac{1}{\sqrt{2}}\begin{pmatrix} Y_1 \\ Y_2 \end{pmatrix} \quad \text{and} \quad Y_- = \frac{1}{\sqrt{2}}\begin{pmatrix} Y_3 \\ Y_4 \end{pmatrix}, \quad (4.3)$$

where,

$Y_1 = [u(1)v(2)w(3) + u(2)v(1)w(3) - u(2)v(3)w(1) - u(3)v(2)w(1)]/\sqrt{4}$,

$Y_2 = [u(1)v(2)w(3) + 2u(1)v(3)w(2) - u(2)v(1)w(3) + u(2)v(3)w(1)$
$\qquad - 2u(3)v(1)w(2) - u(3)v(2)w(1)]/\sqrt{12}$,

$Y_3 = [-u(1)v(2)w(3) + 2u(1)v(3)w(2) - u(2)v(1)w(3) - u(2)v(3)w(1)$
$\qquad + 2u(3)v(1)w(2) - u(3)v(2)w(1)]/\sqrt{12}$

and

$Y_4 = [u(1)v(2)w(3) - u(2)v(1)w(3) - u(2)v(3)w(1) + u(3)v(2)w(1)]/\sqrt{4}$.

One can verify that $P_i \Psi_S = (+1)\Psi_S$, $P_i \Psi_A = (\pm 1)\Psi_A$ and that $P_i Y_\pm = M(P_i)Y_\pm$ where $M(P_i)$ are irreducible (2x2) matrices given by

$$M(P_1) = \begin{pmatrix} 1 & 0 \\ 0 & 1 \end{pmatrix}, \quad M(P_2) = \frac{1}{2}\begin{pmatrix} 1 & -\sqrt{3} \\ -\sqrt{3} & -1 \end{pmatrix}, \quad M(P_3) = \begin{pmatrix} -1 & 0 \\ 0 & 1 \end{pmatrix}$$

(4.4)

$$M(P_4) = \frac{1}{2}\begin{pmatrix} 1 & \sqrt{3} \\ \sqrt{3} & 1 \end{pmatrix}, \quad M(P_5) = \frac{1}{2}\begin{pmatrix} -1 & -\sqrt{3} \\ \sqrt{3} & -1 \end{pmatrix} \quad \text{e} \quad M(P_6) = \frac{1}{2}\begin{pmatrix} -1 & \sqrt{3} \\ -\sqrt{3} & -1 \end{pmatrix},$$

where the permutations $P_i$ are given by $P_1 = (123) \rightarrow (123)$, $P_2 = (123) \rightarrow (213)$, $P_3 = (123) \rightarrow (312)$, $P_4 = (123) \rightarrow (213)$, $P_5 = (123) \rightarrow (132)$ and $P_5 = (123) \rightarrow (321)$.

As well known [46], there are an infinite number of representations of a given group. The above (4.4) irreducible representations of the $S_3$ have been obtained using multiplicative properties of the permutations $P_i$. Another irreducible representation of $S_3$ is obtained, for instance, taking into account the rotations of the equilateral triangle in the Euclidean space $E_3$. To show this we assume that in $E_3$ the states u, v and w occupy the vertices of the equilateral triangle taken in the plane (x,z), as seen in Fig.2. The unit vectors along the axis x, y and z are indicated $\boldsymbol{i}, \boldsymbol{j}$ and $\boldsymbol{k}$, respectively. In Fig.1 the unitary vectors $\mathbf{m}_4$, $\mathbf{m}_5$ e $\mathbf{m}_6$ are given by $\mathbf{m}_4 = -\boldsymbol{k}$, $\mathbf{m}_5 = -(\sqrt{3}/2)\boldsymbol{i} + (1/2)\boldsymbol{k}$ and $\mathbf{m}_6 = (\sqrt{3}/2)\boldsymbol{i} + (1/2)\boldsymbol{k}$, respectively. Representing by Y(123) the initial state with the particles 1, 2 and 3 occupying the vertices u, v and w, respectively, we can show[12,14] that the matrices $M(P_i)$ associated with the permutations $P_i Y = M(P_i)Y$ can be represented by the unitary operators

$$U = \exp[i\,\boldsymbol{j}.\boldsymbol{\sigma}(\theta/2)] \quad\quad e \quad\quad V = i\,\exp[i\,\mathbf{m}_i.\boldsymbol{\sigma}(\varphi/2)], \quad\quad (4.5)$$

where $\boldsymbol{\sigma}$ are Pauli matrices, $\theta = \pm 2\pi/3$ are the rotation angles around the vector $\boldsymbol{j}$, $\varphi = \pm \pi$ are the rotation angles around the vectors $\mathbf{m}_i = \mathbf{m}_4$, $\mathbf{m}_5$ and $\mathbf{m}_6$

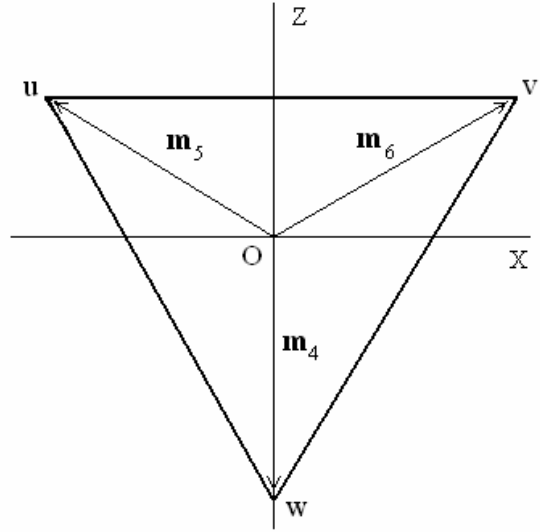

**Figure 2**. The equilateral triangle in the Euclidean space (x,y,z) with vertices occupied by the states u, v and w.

From (4.5) we see that the wavefunctions $\Psi([2,1]) = Y$ are spinors in the Hilbert space $L_2(\varepsilon^{(N)})$ and that when two particles are permuted the functions Y are transformed according to the unitary operators U and V:

$$Y' = P_i Y = UY = \exp[i\,\boldsymbol{j}.\boldsymbol{\sigma}(\theta/2)]Y \quad \text{and} \quad Y' = P_i Y = VY = i\,\exp[i\,\mathbf{m}_i.\boldsymbol{\sigma}(\varphi/2)]Y. \quad (4.6)$$

Assuming that quarks are gentileons with spin ½ and comparing the eigenstates u, v and w with the SU(3)$_c$ color eigenstates b(blue), r(red) e g(green) we would have, instead of the Euclidean axis x and y, the axis $I_3$ e $Y/2$, respectively, of color isospin and hypercharge . In this way the particles permutations would be represented by rotations in the SU(3)$_c$ color space.  In these conditions we verify [12-15] that the color charge $Q_c$ of a barion is a constant of motion which would imply in the quark confinement of the quarks. This constant of motion is a consequence of a global symmetry property of the $S_3$. So, we would expect that if quarks are gentileons they would be permanently confined in hadrons. This confinement would not be obeyed when, for instance, we have a symmetry break of the SU(3)$_c$ group. An approximate gentileonic quantum chromodynamics for hadrons was proposed by Cattani [14] and a model using the Dirac´s equation was elaborated [15] to explain the confinement of gentileonic quarks.

It can easily verified using Eqs.(4.3) that Y = Ψ([2,1]) = 0 when 2 or 3 particles se occupy the same state (u,v,w). In the general case of N particles [8,10,11] the statefunction Ψ of the system is equal to zero only when N − 1 or N particles occupy the same quantum state. We can see that [10,11] for a system with N particles, bosonic, fermionic or gentileonic we have $a_s^* \Psi = (N_s + 1)^{1/2} \Psi$ , $a_s\Psi = \sqrt{N_s} \Psi$ $a_s^* a_s = N_s$ where $N_s$ is the number of particles in the $\varphi_s$ state and $N_s$ is the occupation number operator for the $\varphi_s$ state. For *bosons* are valid the bilinear commutation relations $[a_r, a_s] = [a_r^*, a_s^*] = 0$ e $[a_r^*, a_s] = \delta_{rs}$
For *fermions* are valid the bilinear anticommutation relations $\{a_r, a_s\} = \{a_r^*, a_s^*\} = 0$ and $\{a_r^*, a_s\} = \delta_{rs}$. For the a gentilionic system with 3 particles [10] the operators $a_s^*$ and $a_s$ obey bilinear bosonic relations for the states $Y_1$ and $Y_3$ and obey bilinear fermionic relations for the states $Y_2$ and $Y_4$ Thus, as the gentileonic system is represented by $Y_+$ or $Y_-$, given by (4.3), we see that the Y state is function half fermionic and half bosonic. That is, gentileons are intermediate particles between *bosons* and *fermions*. For 3 gentileons the operators $a_s^*$ and $a_s$ besides the fermionic and bosonic bilinear relations obey trilinear matricial commutation relations [10,14] like, for instance,

$$a_\alpha^* a_\beta^* a_\gamma^* I = M(\alpha\ \beta\ \gamma \rightarrow \gamma\ \beta\ \alpha)\ a_\gamma^* a_\beta^* a_\alpha^* I,$$

$$a_\alpha\ a_\beta^*\ a_\gamma\ I = M(\alpha\ \beta\ \gamma \rightarrow \gamma\ \alpha\ \beta)\ a_\beta^*\ a_\gamma\ a_\alpha I , \qquad (4.7)$$

$$a_\alpha\ a_\beta\ a_\gamma\ I = M(\alpha\ \beta\ \gamma \rightarrow \beta\ \gamma\ \alpha)\ a_\gamma\ a_\alpha\ a_\beta I , \text{ etc...},$$

where the matrices M are given by (4.4) and  I is the unitary (2x2) matrix. It is important to observe that the gentileonic trilinear commutation relations (4.7) are rigorously deduced using the symmetric group $S_3$ and that the trilinear Green´s relations [4] are "anzatz"[10].

Analogously [10] in the general case of systems with N gentileons which is represented by the state Y given by (4.2), the operators $a_s^*$ and $a_s$, for the functions $Y_1$, $Y_3$,..., with odd indices, obey bosonic bilinear commutation relations and for $Y_2$, $Y_4$,..., with even indices, obey fermionic bilinear commutation relations. This shows the hybrid bosonic-fermionic character of the gentileons. Besides these bilinear, bosonic and fermionic relations, the operators $a_s$ and $a_s^*$ obey multilinear matricial commutation relations [10, 14]. These multilinear relations are rigorously deduced using the $S_N$ group.

Due to the complex multilinear matricial commutation relation of the operators $a_s$ e $a_s^*$ and the hybrid nature of the of the gentileonic states Y it was not possible to obtain an exact QFT for gentileons. In this way the *spin-statistics connection* for *gentileons* is still an open problem, as occurs with the *anyons*.

(4.2) *Comparison between the Intermediate and Gentileonic Statistics.*

In both statistics, for a system with any number N of *fermions* the maximum occupation number by a states is d =1 and for any number N of bosons the maximum occupation number is d = N. In this last case in the thermodynamic limit, that is, when N → ∞, we get d → ∞.

On the other side, we verify [8, 10-14] that the functions Y, given by (4.2), representing gentileonic systems with N particles are equal to zero only when N − 1 or N particles occupy the same state $\varphi_s$. This implies that for gentileonic systems with N particles we have always d = N − 1. This behavior is completely different from that proposed by Gentile for systems with N *intermediate particles* for which all values d = 2, 3, ..., N would be possible. So, only in the thermodynamic limit, that is, when N >> 1 and we can do d ≈ N→ ∞, the *Intermediate* and the *Gentileonic* statistics would give equal predictions. In this extreme limit the behavior of the *intermediate particles* and of the *gentileons* are similar to that of *bosons*. The bosonic behavior of the *intermediate particles* was shown by Gentile in the paper "Sopra il fenomeno della condensazione del gás di Bose-Einstein" published in 1941 [2]. In this paper Gentile uses as a starting point the fundamental equation (1.9) of the *Intermediate Statistics* putting d = N obtaining:

$$N_s = Q_s \{1/(\exp[(\varepsilon_s - \mu)/kT] - 1) - (N+1)/(\exp[(N+1)(\varepsilon_s - \mu)/kT] - 1)\}, \quad (4.8)$$

assuming N as finite number. In this way, studying the gas properties he showed that in the N >> 1 limit the Bose-Einstein condensation phenomenon is obtained.

**(5) Summary.**

Many statistical theories have been proposed showing that it could exist particles different from *bosons* and *fermions*. The first one, the *Intermediate Statistics* developed by Gentile [1-3] in a thermodynamic context, would be obeyed by particles called *intermediate particles*. For these particles the maximum occupation number d of a given quantum state would be a finite number that could assume any integer value between 1 and ∞. Recent works developed also in the thermodynamic approach by Ben-Abraham [20] and by Vieira and Tsallis [21] have confirmed the Gentile's predictions.

Pauli in 1940 [26] using the QFT demonstrated a theorem known as Pauli's Theorem. According to this theorem (see also Lüders and Zumino [27] and Burgoyne [28]) "*bosons* are particles with integer spins and their creation and destruction operators $a_k{}^*$ and $a_k$ must obey bilinear commutation relations" and "*fermions* are particles with half-integer spins and their creation and destruction operators $a_k{}^*$ and $a_k$ must obey bilinear anticommutation relations". According to this theorem, as the particles known at that time had integer or half-integer spins, the *intermediate particles* could not exist.

In 1953 Green[4] generalizing the QFT formalism showed that the particles fields, half-integer spin fields or by integer spin fields, could be quantized in such a way that a finite number of particles could occupy the same quantum state. In this new statistics, named *Parastatistics*, the operators $a_s{}^*$ and $a_s$ obey an infinite number of trilinear commutation relations. So according to this new statistics it could exist particles different from *bosons* and *fermions* that were named *parabosons* and *parafermions* or, generically, *paraparticles.*

In 1977 Leinaas e Myeheim [43] showed that in 2-dim systems it could exist particles that would obey a statistic changing continuously between the bosonic and fermionic. In 1982 Wilczek [5-7] showed, confirming the predictions of Leinaas and Myeheim, showed that would have particles with fractionary spins moving in 2-dim systems. These particles have been called *anyons* and obey a statistics that was named *Fractionary Statistics*. Due to intricate mathematical properties it was not possible to calculate exactly [40,41] the wavefunctions $\Psi$ (in non-relativistic or relativistic QM) of systems with many *anyons* and there is no satisfactory QFT formulation for *anyons*. In this way it was not yet possible to analyze the *spin-statistics connection* for *anyons*. However, since *anyons* are intermediate particles between *bosons* and *fermions* the properties of thermodynamic systems of *anyons* are estimated using state equations obtained by interpolation between bosonic and fermionic state equations [42].

In 1984, Cattani and Fernandes [8-15] showed, using the permutation group theory that for 3-dim non-relativistic systems it could exist particles, different from *bosons* and *fermions*. These particles that were named *gentileons* obey the *Gentileonic Statistics*. The gentileonic systems that are formed by a number $N \geq 3$ particles are represented by intermediate Young shapes and their wavefunctions indicated by $\Psi = Y(1,2,..,N)$, seen in Eq.(4.2), are spinors in a Hilbert space $L_2(\varepsilon^{(N)})$ generated by the N! permutations of the particles. To the permutations $P_i$ (i = 1,2,…, N!) are associated the unitary operators $U(P_i)$ that are represented by unitary matrices $M(P_i)$ (or $D(Pi)$) in $L_2(\varepsilon^{(N)})$. For *bosons* and *fermions* these matrices are 1-dim, that is, $P_i \Psi_S = (+1) \Psi_S$ and $P_i \Psi_A = (\pm 1) \Psi_A$. For *gentileons* we have $P_i Y = M(P_i)Y$ where $M(P_i)$ are unitary square matrices (p x p) where $p = N - 1$. For the case of N = 3 the matrices $M(Pi) = U$ and $M(Pi) = V$ can be written as functions of the Pauli matrices: $U = \exp[i \, \mathbf{j}.\boldsymbol{\sigma}(\theta/2)]$ and $V = i \, \exp[i \, \mathbf{m_i}.\boldsymbol{\sigma}(\varphi/2)]$ according to (4.5) that describe the rotations of an equilateral triangle in the Euclidean space $E_3$. For *gentileons*, like for *bosons* and *fermions*, are valid the following relations for the operators $a_s^*$ and $a_s$ : $a_s Y = (N_s + 1)^{1/2} Y$, $a_s^* Y = \sqrt{N_s} \, Y$, $a_s^* a_s = N_s$ where $N_s$ is the occupation number operator of the s state and $N_s$ is the number of particles in the s state. According to (4.2) the statefunction Y of a gentileonic system with N particles is described by $\tau$ base vectors $Y_1, Y_{2,....,} Y_\tau$. The functions with odd índices $Y_1, Y_3, ...$, show a bosonic behavior, that is, they obey bilinear commutation relations $[a_r, a_s] = [a_r^*, a_s^*] = 0$ and $[a_r^*, a_s] = \delta_{rs}$ On the other hand, the functions with even indices $Y_2, Y_4,..$, show a fermionic behavior, that is, they obey bilinear anticommutation relations $\{a_r, a_s\} = \{a_r^*, a_s^*\} = 0$ and $\{a_r^*, a_s\} = \delta_{rs}$. This implies that the eigenfunctions Y have a hybrid bosonic-fermionic nature. So, *gentileons* are intermediate particles between bosons and fermions as occurs with *anyons*. In addition, the operators $a_s^*$ and $a_s$ obey multilinear matricial commutation relations. Due to these intricate properties it was not possible to obtain an exact QFT for *gentileons*. Consequently, the *spin-statistics connection* for gentileons is still an open problem. Finally, there is an essential difference between the *Intermediate* and *Gentileonic* statistics: according to Gentile [1-3] it could exist many systems composed by N intermediate particles which would have different maximum occupation numbers d = 2, 3, ..., N. However, following Cattani and Fernandes [8-15] it is not possible since a system with N gentileons can only have $d = N - 1$. The permutation group theory excludes the other possibilities.

**(6) Conclusions and Discussions.**

Gentile used the *Intermediate Statistics* to analyze the Bose-Einstein condensation [2] and the liquid helium superfluidity [3]. Other works using this statistics to study properties of thermodynamic systems were done, for instance, by Shelomo [20] and by Vieira e Tsallis [21]. From these works we can conclude that up to date all known thermodynamic processes can be explained using simply the bosonic, fermionic and Maxwell-Boltzmann statistics. There is no observed thermodynamic system or process that could eventually give support to the existence of the *intermediate particles*.

The Parastatistics is still an open and vast field of researches. Many works were published and are being published on the subject [29-33]. The fundamental result of this statistics was to show that particles with integer or half-integer spins are not necessarily *bosons* or *fermions*, respectively. An aspect that we consider unsatisfactory: the field $\Psi(x)$ of the *paraparticles* is not able to describe in the non-relativistic limit [29] the gentileonic states Y as we ought to expect [10,49].

The existence of anyons, predicted by the Fractionary Statistics, that are quasiparticles with fractionary spins was confirmed in the context of the Fractionary Quantum Hall effect. The *anyons* detected in this effect, that were named "Laughlin Particles", are quasiparticles with spin and charge fractionary [3-37].

From the *Parastatistics* and Fractionary *Statistics* predictions result two fundamental properties: (a) particles in 3-dim systems with integer and half-integer spins are not necessarily *bosons* or *fermion*s, respectively. (b) Exist particles in 2-dim systems with spins that can assume any value between integer and half-integer numbers.

Finally let us consider the *Gentileonic Statistics* which is valid to 3-dim non-relativistic systems. The main characteristics predicted to the gentileons are that they are permanently confined and that their systems cannot coalesce. These properties suggest that they are *quasiparticles*. Taking into account the item (a) mentioned above we see that they could have integer or half-integer spins. As no exact QFT was developed for gentileons it was not possible to study for them the *spin-statistics connection*. Consequently we can only make conjectures about their spin values. Trying to explain the quark confinement and the non-coalescence of the hadrons we have assumed [9-14] that quarks could be gentileons with spin ½. With this hypothesis and using only the symmetry properties of the group $S_N$ it was possible to explain satisfactorily essential aspects of the quark confinement and the non-coalescence of hadrons. An approximate quantum chromodynamics was proposed [14] for gentileonic hadrons. A simple model using the gauge theory and the Dirac's equation was also constructed [15] to explain the quark confinement.

Taking into account the large number of works done about conventional and non-conventional Statistics there is a strong tendency to believe that the *real particles* can only be *bosons* and *fermions* and that all other particles, different from *bosons* and *fermions*, would be *quasiparticles*. Up to date in 3-dim systems only *bosons* and *fermions* have been detected. In 2-dim systems have been detected the *quasiparticles* called anyons, which have charge and spin fractionary. Would it be possible that quarks, which have fractionary charges, are *quasiparticles*? If it is true they would be entities permanently confined as occurs with *gentileons*.

REFERENCES
[1]G.Gentile Jr. Nuovo Cimento 1, 493 (1940).
[2]G.Gentile Jr. Ric.Sci. 12, 341 (1941).
[3]G.Gentile Jr. Nuovo Cimento 19, 109 (1942).
[4]H.S.Green. Phys.Rev. 90, 270 (1953).
[5]F. Wilczek, Phys.Rev.Lett.48, 1144 (1982).
[6]F. Wilczek. Phys.Rev.Lett.49, 957 (1982).
[7]F.Wilczek "Fractional Statistics and Anyon Superconductivity".World Scientific (1990).
[8]M. Cattani and N. C. Fernandes. Rev. Bras. Fís.12,585(1982).
[9]M. Cattani and N. C. Fernandes. Rev. Bras. Fís. 13, 464 (1983)
[10]M. Cattani and N.C.Fernandes. Nuovo.Cimento 79A, 107 (1984).
[11]M. Cattani and N.C.Fernandes. Nuovo.Cimento 87B, 70 (1985).
[12]M. Cattani and N.C.Fernandes. Phys. Lett. 124A, 229 (1987),
[13]M. Cattani and N.C.Fernandes. An. Acad. Bras. de Ciências 59, 283(1987).
[14]M. Cattani. Acta Physica Polonica B20, 983 (1989).
[15]M. Cattani. An. Acad. Bras. de Ciências 67 (2), 153 (1995).
[16]A.Sommerfeld. "Thermodynamics and Statistical Mechanics", Academic Press (1964).
[17]L. D. Landau e E. M. Lifshitz. "Statistical Mechanics", Pergamon Press Ltd. (1958).
[18]K. Huang."Statistical Mechanics", John Wiley & Sons (1963).
[19]F. W. Sears. "Thermodynamics", Addison Wesley & Sons (1953).
[20]S. I .Ben-Abraham. Am. J. Phys. 38, 1335 (1970).
[21]M.C. de Souza Vieira and C. Tsallis. Journal of Statistical Physics 48, 97(1987).
[22]P. A. M. Dirac. "The Principles of Quantum Mechanics", Oxford Press (1947).
[23]L. Landau and E. Lifchitz. "Mécanique Quantique", Éditions Mir (1966).
[24]L.I.Schiff . "Quantum Mechanics" , McGraw-Hill (1955).
[25]P.Roman. "Theory of Elementary Particles", North - Holland (1964).
[26]W.Pauli. Phys. Rev. 58, 716 (1940).
[27]G.Lüders and B.Zumino. Phys. Rev.110, 1450 (1958).
[28]N.Burgoyne. Nuovo Cimento 8, 607 (1958).
[29]Y.Ohnuki and S.Kamefuchi."Quantum Field Theory and Parastatistics".Springer-Verlag (1990).
[30]O.W.Greenberg and A. K. Mishra, Phys. Rev. D70, 125013 (2004).
[31]M.V.Mededev, Phys. Rev. Lett. 78, 4147 (1997).
[32]S. Chaturvedi, Phys. Rev. E54, 1378 (1996).
[33]M. D.Gould and J.Paldus, Phys.A34, 804 (1986).
[34]O.W. Greenberg, Phys. Rev. Lett., 598 (1964).
[35]F. E. Camino, W.Zhou and V. J. Goldman, Phys. Rev B72, 075342 (2005).
[36]J. M. F. Bassalo – www.searadaciencia.ufc.br/curiosidadesdafisica.
[37]D.Lindley, Phys. Rev. Focus, in http://focus.aps.org/story/v16/st14 in June 2007.
[38]Z.N.C.Ha, "Quantum-Many Body Systems in One Dimension".World Scientific(1996).
[39]J. K. Jain, Phys. Rev. Lett. 63, 199 (1989).
[40]S. Forte, Rev. Mod. Phys. 64 , 193 (1992).
[41]D. P.Arovas, R.Schrieffer and F. Wilczek . N. Phys. B251, 117(1985). ]
[42]S.Viefers, F.Ravndal and T.Haugset. Am. J. Phys. 63, 369 (1995).
[43]J.M.Leinaas and J.Myrheim. Nuovo Cimento B37, 1 (1977).
[44]R. Mills. Am. J. Phys .57, 493(1989).
[45]K. Moriyasu."An Elementary Primer for Gauge Theory", World Scientific (1983).

[46]J.M.F. Bassalo and M. Cattani. "Teoria de Grupos", Livraria da Física (2008).
[47]H.Erlichson, Am. J. Phys. 38, 162 (1970).
[48]T.T.Wu and C.N.Yang. Phys. Rev.D12, 3845 (1975).
[49]T.Okayama. Prog.Theor.Phys. 7, 517 (1952).